

\magnification 1200
\baselineskip=12pt
\hsize=15.3truecm \vsize=22 truecm \hoffset=.1truecm
\parskip=14 pt
\overfullrule=0pt

\catcode`@=11
\def\ifundefined#1{\expandafter\ifx\csname
                        \expandafter\eat\string#1\endcsname\relax}
\def\atdef#1{\expandafter\def\csname #1\endcsname}
\def\atedef#1{\expandafter\edef\csname #1\endcsname}
\def\atname#1{\csname #1\endcsname}
\def\ifempty#1{\ifx\@mp#1\@mp}
\def\ifatundef#1#2#3{\expandafter\ifx\csname#1\endcsname\relax
                                  #2\else#3\fi}
\def\eat#1{}
\newcount\refno \refno=1
\def\labref #1 #2 #3\par{\atdef{R@#2}{#1}}
\def\lstref #1 #2 #3\par{\atedef{R@#2}{\number\refno}
                              \advance\refno by1}
\def\txtref #1 #2 #3\par{\atdef{R@#2}{\number\refno
      \global\atedef{R@#2}{\number\refno}\global\advance\refno by1}}
\def\doref  #1 #2 #3\par{{\refno=0
     \vbox {\everyref \item {\reflistitem{\atname{R@#2}}}
            {\d@more#3\more\@ut\par}\par}}\refskip }
\def\d@more #1\more#2\par
   {{#1\more}\ifx\@ut#2\else\d@more#2\par\fi}
\def\@cite #1,#2\@ver
   {\eachcite{#1}\ifx\@ut#2\else,\hskip0pt\@cite#2\@ver\fi}
\def\cite#1{\citeform{\@cite#1,\@ut\@ver}}
\def\eachcite#1{\ifatundef{R@#1}{{\tt#1??}}{\atname{R@#1}}}
\def\defonereftag#1=#2,{\atdef{R@#1}{#2}}
\def\defreftags#1, {\ifx\relax#1\relax \let\next\relax \else
           \expandafter\defonereftag#1,\let\next\defreftags\fi\next }
\def\@utfirst #1,#2\@ver
   {\author#1,\ifx\@ut#2\afteraut\else\@utsecond#2\@ver\fi}
\def\@utsecond #1,#2\@ver
   {\ifx\@ut#2\andone\author#1,\afterauts\else
      ,\author#1,\@utmore#2\@ver\fi}
\def\@utmore #1,#2\@ver
   {\ifx\@ut#2\and\author#1,\afterauts\else
      ,\author#1,\@utmore#2\@ver\fi}
\def\authors#1{\@utfirst#1,\@ut\@ver}
\catcode`@=12
\let\REF\labref
\def\citeform#1{{\bf\lbrack#1\rbrack}}  
\let\reflistitem\citeform               
\let\everyref\relax                     
\let\more\relax                         
\def\refskip{\vskip 10pt plus 2pt}      
\def\colbr{\hskip.3em plus.3em\penalty-100}  
\def\combr{\hskip.3em plus4em\penalty-100}   
\def\refsecpars{\emergencystretch=50 pt      
                 \hyphenpenalty=100}
\def\Bref#1 "#2"#3\more{\authors{#1}:\colbr {\it #2},\combr #3\more}
\def\Gref#1 "#2"#3\more{\authors{#1}\ifempty{#2}\else:\colbr``#2''\fi
                        ,\combr#3\more}
\def\Jref#1 "#2"#3\more{\authors{#1}:\colbr``#2'',\combr \Jn#3\more}
\def\inPr#1 "#2"#3\more{in: \authors{\eds#1}:\colbr
                          ``{\it #2}'',\combr #3\more}
\def\Jn #1 @#2(#3)#4\more{\hbox{\it#1}\ {\bf#2}(#3)#4\more}
\def\author#1. #2,{\hbox{#1.~#2}}            
\def\sameauthor#1{\leavevmode$\underline{\hbox to 25pt{}}$}
\def\and{, and}   \def\andone{ and}          
\def\noinitial#1{\ignorespaces}
\let\afteraut\relax
\let\afterauts\relax
\def\etal{\def\afteraut{, et.al.}\let\afterauts\afteraut
           \let\and,}
\def\eds{\def\afteraut{(ed.)}\def\afterauts{(eds.)}}
\catcode`@=11
\newcount\eqNo \eqNo=0
\def\lasteq{\secNo.\number\eqNo}
\def\deq#1(#2){{\ifempty{#1}\global\advance\eqNo by1
       \edef\n@@{\lasteq}\else\edef\n@@{#1}\fi
       \ifempty{#2}\else\global\atedef{E@#2}{\n@@}\fi\n@@}}
\def\eq#1(#2){\edef\n@@{#1}\ifempty{#2}\else
       \ifatundef{E@#2}{\global\atedef{E@#2}{#1}}%
                       {\edef\n@@{\atname{E@#2}}}\fi
       {\rm($\n@@$)}}
\def\deqno#1(#2){\eqno(\deq#1(#2))}
\def\deqal#1(#2){(\deq#1(#2))}
\def\eqback#1{{(\advance\eqNo by -#1 \lasteq)}}

\def\eqgroup(#1){{\global\advance\eqNo by1
       \edef\n@@{\lasteq}\global\atedef{E@#1}{\n@@}}}
\outer\def\iproclaim#1/#2/#3. {\vskip0pt plus50pt \par\noindent
     {\bf\dpcl#1/#2/ #3.\ }\begingroup \interlinepenalty=250\lessblank\sl}
\newcount\pcNo  \pcNo=0
\def\lastpc{\number\pcNo} 
\def\dpcl#1/#2/{\ifempty{#1}\global\advance\pcNo by1
       \edef\n@@{\lastpc}\else\edef\n@@{#1}\fi
       \ifempty{#2}\else\global\atedef{P@#2}{\n@@}\fi\n@@}
\def\pcl#1/#2/{\edef\n@@{#1}%
       \ifempty{#2}\else
       \ifatundef{P@#2}{\global\atedef{P@#2}{#1}}%
                       {\edef\n@@{\atname{P@#2}}}\fi
       \n@@}
\def\Def#1/#2/{Definition~\pcl#1/#2/}
\def\Thm#1/#2/{Theorem~\pcl#1/#2/}
\def\Lem#1/#2/{Lemma~\pcl#1/#2/}
\def\Prp#1/#2/{Proposition~\pcl#1/#2/}
\def\Cor#1/#2/{Corollary~\pcl#1/#2/}
\def\Exa#1/#2/{Example~\pcl#1/#2/}
\font\sectfont=cmbx10 scaled \magstep2
\def\bgsecti@n #1. #2\e@h{\def\secNo{#1}\eqNo=0}
\def\bgssecti@n#1. #2\e@h{}
\def\secNo{00}
\def\lookahead#1#2{\vskip\z@ plus#1\penalty-250
  \vskip\z@ plus-#1\bigskip\vskip\parskip
  {#2}\nobreak\smallskip\noindent}
\def\secthead#1. #2\e@h{\leftline{\sectfont
                        \ifx\n@#1\n@\else#1.\ \fi#2}}
\def\bgsection#1. #2\par{\bgsecti@n#1. #2\e@h
        \lookahead{.3\vsize}{\secthead#1. #2\e@h}}
\def\bgssection#1. #2\par{\bgssecti@n#1. #2\e@h
        \lookahead{.3\vsize}{\leftline{\bf#1. #2}}}
\def\bgsections#1. #2\bgssection#3. #4\par{%
        \bgsecti@n#1. #2\e@h\bgssecti@n#3. #4\e@h
        \lookahead{.3\vsize}{\vtop{\secthead#1. #2\e@h\vskip10pt
                             \leftline{\bf#3. #4}}}}
\def\Acknow#1\par{\ifx\REF\doref
     \bgsection. Acknowledgements\par#1\refsecpars
     \bgsection. References\par\fi}
\catcode`@=12
\def\class#1 #2*{{#1},}
\def\nl{\hfill\break}

\def\idty{{\leavevmode{\rm 1\ifmmode\mkern -5.4mu\else
                                            \kern -.3em\fi I}}}
\def\Ibb #1{ {\rm I\ifmmode\mkern -3.6mu\else\kern -.2em\fi#1}}
\def\Ird{{\hbox{\kern2pt\vbox{\hrule height0pt depth.4pt width5.7pt
    \hbox{\kern-1pt\sevensy\char"36\kern2pt\char"36} \vskip-.2pt
    \hrule height.4pt depth0pt width6pt}}}}
\def\Irs{{\hbox{\kern2pt\vbox{\hrule height0pt depth.34pt width5pt
       \hbox{\kern-1pt\fivesy\char"36\kern1.6pt\char"36} \vskip -.1pt
       \hrule height .34 pt depth 0pt width 5.1 pt}}}}

\def\ibbt #1{\leavevmode\hbox{\kern.3em\vrule
     height 1.5ex depth -.1ex width .2pt\kern-.3em\rm#1}}
\def\ibbs#1{\hbox{\kern.25em\vrule
     height 1ex depth -.1ex width .2pt
                   \kern-.25em$\scriptstyle\rm#1$}}
\def\ibbss#1{\hbox{\kern.22em\vrule
     height .7ex depth -.1ex width .2pt
                   \kern-.22em$\scriptscriptstyle\rm#1$}}

  \def\Rl {{\Ibb R}}

\def\lessblank{\parskip=5pt \abovedisplayskip=2pt
          \belowdisplayskip=2pt }

\def\proof#1{\par\noindent {\bf Proof #1}\          
         \begingroup\lessblank\parindent=0pt}
\def\QED {\hfill\endgroup\break
     \line{\hfill{\vrule height 1.8ex width 1.8ex }\quad}
      \vskip 0pt plus100pt}
\ifundefined\Sec\def\Sec#1{Section~{#1}}\fi
\newcount\figcount  \figcount=1
\def\Fig#1{Figure~\ifatundef{F@#1}%
                  {{\tt?\message{Figure #1 undefined}}}%
                  {\atname{F@#1}}}
\def\defoneFig#1,{\atedef{F@#1}{\number\figcount}%
                   \global\advance\figcount by 1}
\def\defFigs#1, {\ifx\relax#1\relax \let\next\relax \else
           \expandafter\defoneFig#1,\let\next\defFigs\fi\next }
\def\onlycap#1\ad#2(#3@#4)#5\\{\vfill
     \centerline{\hbox to 12truecm{\hsize12truecm\parindent=0pt
      \vtop{{\it\Fig{#1}\ifx@#2@\else\ (for {#2})\fi:}#5}}} }

\def\Exa#1{Example~\ifatundef{X@#1}%
                  {\tt?\message{Example #1 undefined}}%
                  {\atname{X@#1}}}
\def\defoneExa#1,{\atedef{X@#1}{\number\figcount}%
                   \global\advance\figcount by 1}
\def\defExas#1, {\ifx\relax#1\relax \let\next\relax \else
           \expandafter\defoneExa#1,\let\next\defExas\fi\next }
\def\EXA #1:#2: #3\par{\par\vskip\parskip\noindent{\bf\Exa{#1}:\
                      #2\hfill\vskip-10pt\noindent
                  {\it#3}\par\noindent}}


\def\conv{\mathop{\rm conv}\nolimits}
\def\Exp#1{e^{\textstyle#1}}
\def\Order{{\bf O}}
\def\Set#1#2{#1\lbrace#2#1\rbrace}  
\def\abs #1{{\left\vert#1\right\vert}}
\def\ad#1{{\rm ad}\,#1}
\def\ad{{\rm ad}}
\def\bra #1>{\langle #1\rangle}
\def\bracks #1{\lbrack #1\rbrack}

\def\ket #1>{\mid#1\rangle}

\def\norm #1{\left\Vert #1\right\Vert}

\def\ot{\otimes}
\def\rstr{\hbox{$\vert\mkern-4.8mu\hbox{\rm\`{}}\mkern-3mu$}}
\def\set #1{\left\lbrace#1\right\rbrace}

\def\stt{\,\vrule\ }
\def\th{\hbox{${}^{{\rm th}}$}\ }  

\def\tr{\mathop{\rm Tr}\nolimits}
\def\trexp{\tr\,\exp}
\font\bigf=cmr10 scaled \magstep 2
\def\Tr#1{\lower2pt\hbox{\kern.2em{\bigf Tr}\lower5pt\hbox{$
          \scriptstyle #1$}}\;}

\def\subsetnoteq{\lower4pt\hbox{$\buildrel\subset\over\neq$}}
\def\smallpmatrix#1{\null\left(\vcenter{\normalbaselines\mathsurround=0pt
    \ialign{\hfil$\scriptstyle##$\hfil&&\ \hfil$\scriptstyle##$\hfil\crcr
      \mathstrut\crcr\noalign{\kern-\baselineskip}
      #1\crcr\mathstrut\crcr\noalign{\kern-\baselineskip}}}\right)}
\def\lntw{\widetilde{\ln}\,}
\def\oneten{\idty\otimes}

\def\class#1 #2*{{#1},}

\def\phi{\varphi}            
\def\epsilon{\varepsilon}    
\def\A{{\cal A}} \def\B{{\cal B}} \def\C{{\cal C}} \def\D{{\cal D}}
   
\def\M{{\cal M}}  
   \def\Z{{\cal Z}}

\def\Htot{H_{\rm tot}}
\def\Ybd{Y_{\rm bd}}
\def\omin{\om_{\rm int}}
\def\muin{\mu_{\rm int}}
\def\Ain{\A_{\rm int}}
\def\Abd{{\cal A}_{\rm bd}}
\def\Aout{\A_{\rm ext}}
\def\condex#1{{\Ibb E}^{\mkern-2mu#1}} 
\def\Cc#1{\C(\set{1,\ifx#122\else\if#132,3\else\ldots,#1\fi\fi})}
\def\RE{_{\rm RE}}
\def\AH{_{\rm AH}}
\def\EE{_{\rm EE}}
\def\SAH{{\cal E}_{\rm AH}}
\def\SAHp{{\cal E}_{\rm AH}^{\rm pure}}
\def\SRE{{\cal E}_{\rm R}}
\def\GAH{{\cal G}_{\rm AH}}
\def\GAHp{{\cal G}_{\rm AH}^{\rm pure}}
\def\SRG{{\cal G}_{\rm R}}
\def\SEE{{\cal E}_{\rm EE}}

\def\ie{i.e., }               
\def\eg{e.g., }               
\def\om{\omega}

\defFigs C4C3, Q2Q2, C3Q2a, C3Q2b, C3C2, , \figcount=1
\defExas C4C3, C2Qn, Q2Q2, QCQ, C3Q2, C3C2, CnC3, ,
\let\figures0
\defreftags GoldThom=Ara, GibbsCond=AI, BMV=BMV, BraRo=BR,
Dobrushin=DS, FCS=FNW, EnEnt=FV, TXP=FW, Freuden=FF, Gaudi=Gau,
Gerisch=Ger, FGH=GW, Israel=Isr, Miek=Mie, Pasquier=PS, RelEnt=RW,
Sewell=Sew, Vets=Vets, Mathematica=Mat, ,

\font\BF=cmbx10 scaled \magstep 3
\line{\tt cond-mat/9504090\hfill Preprint KUL-TF-95/11}
\voffset=2\baselineskip
\hrule height 0pt
\vskip 40pt plus40pt
\centerline{{\BF Boundary Conditions for}}
\vskip10pt
\centerline{{\BF  Quantum Lattice Systems}}
\vskip 30pt plus30pt
\centerline{
  M.~Fannes$^{1,2}$ and R.F.~Werner$^3$}
\vskip 20pt plus40pt

\noindent {\bf Abstract}\hfill\break
For classical lattice systems, the Dobrushin-Lanford-Ruelle theory
of boundary conditions states that the restriction of a global
equilibrium state to a subsystem can be obtained as an integral over
equilibrium states of the subsystem alone. The Hamiltonians for the
subsystem are obtained by fixing a configuration for the variables
in the complement of the subsystem, or more generally, by evaluating
the full interaction Hamiltonian with respect to a state for the
complement. We provide examples showing that the quantum mechanical
version of this statement is false. It fails even if the subsystem
is classical, but embedded into a quantum environment. We suggest
an alternative characterization of the local restrictions of global
equilibrium states by inequalities involving only local data.

\noindent {\bf Mathematics Subject Classification (1991):}
\hfill\break
\class 82B10   Quantum equilibrium statistical mechanics
                   (general)*
\class 46L60   Applications of selfadjoint operator algebras to
               physics*
\class 82B20   Lattice systems (Ising, dimer, Potts, etc.)*

\def\Email{\nl\vrule width0pt\qquad Email: \tt }
\vfootnote1
  {Inst. Theor. Fysica, Universiteit Leuven, B-3001 Heverlee,
   Belgium.
  \Email mark.fannes@fys.kuleuven.ac.be }
\vfootnote2
  {Onderzoeksleider, NFWO Belgium}
\vfootnote3
  {Fachbereich Physik, Universit\"at Osnabr\"uck,
  49069 Osnabr\"uck, Germany.
  \Email reinwer@dosuni1.rz.uni-osnabrueck.de}
\vfill\eject

\bgsection 1. Introduction

It is well-known that basic phenomena in statistical physics such as
phase transitions, critical behaviour, and symmetry breaking cannot
be modelled by finite systems as studied in classical or quantum
mechanics. On the other hand, infinite systems can rarely be handled
directly by finite computations. Therefore it is a fundamental task
of statistical mechanics to develop techniques by which properties
of infinite systems can be inferred from a study of finite
subsystems. For the equilibrium statistical mechanics of classical
systems on a lattice there is a standard tool for performing
precisely such a step, namely the {\it Dobrushin-Lanford-Ruelle}
(DLR)~theory of boundary conditions~\cite{Israel}. It says that the
effect of the ambient infinite system on a finite subsystem can be
parametrized adequately by a choice of ``boundary conditions'' for
the finite subsystem. Here, boundary conditions are identified with
configurations of the particles outside the finite subsystem in
consideration. They can then be used to classify the pure phases, or
to select a particular pure phase in the event of a broken symmetry.
In specific models such as, \eg the two-dimensional Ising model, a
small selection of boundary conditions suffices to obtain all
phases.

The notion of boundary condition as a configuration in the boundary
region of a finite subsystem is the starting point both for proofs
of existence of a phase transition and proofs of its opposite, the
uniqueness of the equilibrium state. An example of the first is the
Peierls argument~\cite{Sewell} where one shows that the
configuration of a small subsystem (a single spin), strongly depends
on the configurations of the boundary region. An example of the
other usage is the Dobrushin-Shlosman unicity
argument~\cite{Dobrushin}. In this case, the influence of variations
in boundary conditions is proven to become negligible when one
surrounds a small system by larger and larger intermediate regions.

It is one aim of the present paper to show, by way of explicit
examples that, while the basic statement of the DLR-theory has a
straightforward generalization to the case of quantum lattice
systems, this generalization, along with some of its natural
modifications, turns out to be false. But our intention is not
merely to disprove a single statement. Rather, we believe that a
careful study of examples may be helpful for the development of the
much needed tools of quantum statistical mechanics which would be
capable to serve some of the functions of the classical DLR-theory.
The energy-entropy inequalities \cite{EnEnt} described in \Sec4
are a promising step in this direction: they form a set of strictly
local constraints on the restrictions of possible equilibrium
states, which approaches an exact characterization of equilibrium
states as the local region under consideration becomes larger and
larger. For the restrictions to classical subsystems, an efficient
evaluation of these conditions is possible (see \Sec5), but, for
the case of general quantum systems, better techniques of extracting
information from these inequalities are still to be developed.

The term boundary condition, especially as used in the context of
quantum statistical mechanics, carries several meanings, often not
too clearly defined. ``Free boundary conditions'' refer to a special
way of constructing finite volume Hamiltonians from an interaction
potential, namely as the sum over all potential terms localized in
the given region. There are always many different interaction
potentials representing the same infinite volume Hamiltonian
(``equivalent potentials'' \cite{Israel}), which in this language
lead to different ``free boundary conditions'' for the same
Hamiltonian. Some properties, like the presence or absence of
frustration \cite{Miek,FGH}, or the invariance of local Hamiltonians
under the action of a symmetry group or quantum group
\cite{Pasquier}, depend very sensitively to this choice. ``Periodic
boundary conditions'' arise when a lattice is shaped into a torus.
The translation symmetry of the infinite lattice is then
approximated by cyclic shifts. Another type of ``boundary
condition'' is obtained by adding an overall small external field,
which is removed afterwards \cite{Gerisch}. Such fields are usually
introduced for driving the system in a pure phase of broken
symmetry. Typically, all these additional terms to the Hamiltonian
leave the thermodynamic quantities unchanged. In the description of
continuous systems ``boundary conditions'' are used to specify the
Hamiltonian as a self-adjoint operator \cite{BraRo}. This can be
used to confine the system to a particular region, or to introduce
hard-core repulsion. In principle, this is equivalent to adding
strongly repulsive potentials to the Hamiltonian. Finally,
conditions that select particular ground states in situations with
degeneracy are also called boundary conditions. An example of such
usage occurs in valence-bond-solid models \cite{FCS}.

For simplicity, we will only consider finite systems in this paper.
The difficulties that we demonstrate for describing the local
restrictions of equilibrium states  will certainly not be lifted by
considering infinite systems. (This could only happen if more
structure, like translation invariance, is incorporated into the
scheme). However, the notions that we introduce for finite systems
can be generalized in a straightforward manner to the infinite
systems obtained in the thermodynamical limit. The adequate tools
for doing so are the DLR-equations for classical lattice systems,
the KMS-condition~\cite{BraRo} for quantum spin systems, and the
energy-entropy inequalities for both. The basic input, namely the
interaction between the lattice spins, now translates into a
relative Hamiltonian for the classical systems, and a derivation that
generates the dynamics for quantum spin systems. There is an,
apparently local, description of equilibrium states for quantum spin
systems known as the Gibbs condition~\cite{GibbsCond} (see also
Sect. 6.2 of \cite{BraRo}). Unfortunately, it involves a variation
over all global extensions of the local state, and is therefore of
little practical help.

The paper is organized as follows: in \Sec2 we briefly review
the local ``DLR-characterization'' of equilibrium states of
classical systems in terms of boundary conditions. \Sec3
introduces the corresponding notions for quantum spin systems. In
\Sec4 we recall the characterization of equilibrium states in
terms of energy-entropy inequalities, while \Sec5 examines the
consequences of these inequalities for the restrictions of global
equilibrium states to a classical subsystem. \Sec6 contains the
major part of our results, a succession of examples showing the
failure of the direct quantum analogue of the DLR-theory. Where
feasible we also discuss the consequences of the energy-entropy
inequalities.

\bgsection 2. \vtop{\noindent\baselineskip=20pt
  Boundary conditions for classical \hfill\break
  equilibrium states}

In this section we will only consider finite, discrete classical
systems, described by finite configuration spaces. We will
concentrate on a distinguished subsystem, called the inner system,
with configuration space $X$. The configuration space of the global
system is $X\times Y$, where $Y$ is the configuration space of the
outer system. We assume that we have a complete knowledge of the
contribution of the particles in the inner system to the total
energy $\Htot$, meaning that we write
$$\eqalign{
  \Htot= H+ V\quad, \quad
  &H\ {\rm fixed\ function\ on\ }X\times Y\ {\rm and} \cr
  &V\ {\rm only\ depending\ on\ } Y\quad.}
\deqno(hcc)$$
The function $(x,y)\in X\times Y\mapsto H(x,y)$ is the basic information
that is at our disposal, whereas the size of the outer system and the
interaction $V$ is supposed to be less well-known. The minimal choice of
outer system is fixed by requiring that $H$ actually depends on the
configurations in this minimal exterior region. More precisely we say that
$y$ and $y'$ in $Y$ are equivalent iff for all $x\in X$, $H(x,y)= H(x,y')$.
The quotient of $Y$ by this equivalence relation is called the set of the
boundary configurations $\Ybd$ and it is the minimal exterior configuration
space that has to be considered. We will sometimes refer to it as the
boundary system.

The equilibrium state of the global system at inverse temperature $\beta=
1/k\,T$ is given by the canonical Gibbs distribution, which
assigns the probability
$$\mu(x,y)= \Exp{-\beta \Htot(x,y)}/\Z
\deqno(gibbsm)$$
to the configuration $(x,y)$. The normalization factor is the canonical
partition function $\Z= \sum_{x,y} e^{-\beta \Htot(x,y)}$. Unless we
explicitly discuss temperature behaviour, we assume that $\beta$ is
included in the Hamiltonian, and therefore put $\beta=1$.

The restriction $\muin$ of $\mu$ to $X$ is easily computed. It assigns to
$x\in X$ the probability of $\Set{}{(x,y)\stt y\in Y}$:
$$\eqalignno{
 \muin(x)
 &= \sum_{y\in Y} \mu(x,y)
 = \sum_{y\in Y} {\Exp{-\Htot(x,y)}\over \Z} \cr
 &= \sum_{y\in Y} {\Z^y\,\Exp{-V(y)}\over\Z}\, {\Exp{-H^y(x)}\over \Z^y}
 = \sum_{y\in Y} \lambda(y)\, \mu^y(x)
\quad, &\deqal(dlr)}$$
where
$$\mu^y(x)= \Exp{-H^y(x)} /\Z^y
\deqno(gibbsmr)$$
is the Gibbs distribution on $X$ defined by the averaged Hamiltonian
$$H^y(x)= H(x,y)
\quad, \deqno(ahpc)$$
and where the $\lambda(y)$ are positive numbers adding up to $1$.
Thus formula \eq(dlr) says that the global equilibrium state,
reduced to the inner system $X$, is a convex combination of
equilibrium states $\mu^y$ of the inner subsystem for averaged
Hamiltonians $H^y$. The configurations $y\in\Ybd$ can be viewed as
boundary conditions, defining the averaged Hamiltonians $H^y$.
Formula~\eq(dlr) solves the problem of giving a simple
characterization of the equilibrium distributions, reduced to the
inner system, in terms of the basic interaction $H$.
It is a basic tool in proving both absence and occurrence of phase
transitions.

It is easily seen that all the Gibbs distributions $\mu^y$ are
themselves limits of restrictions of equilibrium distributions of
the global system to $X$. Indeed, it suffices to consider the
minimal system $X\times\Ybd$ and to take $V=\lambda\,e_{y}$ (where
the function $e_{y}$ takes the value $1$ on $y$ and zero
elsewhere), and consider the limit $\lambda\to-\infty$. This makes
the configuration $y\in\Ybd$ infinitely attractive to the outer
system, and yields a sequence of equilibrium states, whose
restrictions converge to $\mu^y$. This procedure is reminiscent of
the way one finds the Dirichlet boundary conditions for the
Schr\"odinger equation in a finite region, by introducing larger and
larger confining potentials.

The set of restrictions of global equilibrium states $\mu$,
determined by Hamiltonians $\Htot$ satisfying~\eq(hcc), is convex.
This is easily seen as follows: if $Y$ and $Y'$ are two outer
configuration spaces for the subsystem, consider then $Y''= Y\cup
Y'$ and the total Hamiltonian $\Htot''= H+ V''$, where $V''(y)=
V(y)$ if $y$ belongs to $Y$ and $V''(y)= V'(y)+ C$ else. By varying
the constant $C\in\Rl$, we can produce any convex combination of the
restrictions of the equilibrium distributions corresponding to
$\Htot$ and $\Htot'$.
Taking the conclusions of the last three paragraphs together, we
arrive at the following statement, whose quantum versions form the
subject of this paper:
\vskip10pt
\vtop{\narrower\parskip=0pt\noindent\it
for a measure $\mu$ on $X$ the following two statements are
equivalent:
\item{(1)} $\mu$ can be approximated by the restriction of
Gibbs distributions for Hamiltonians in the class \eq(hcc).
\item{(2)}
$\mu$ is a convex combination of the Gibbs distributions
$\mu^y$, formed with Hamiltonians $H^y$  \eq(ahpc), averaged with
one outside configuration $y\in\Ybd$. }
\noindent
Especially suggestive in this equivalence is the identification of
the ``extreme'' elements $\mu^y$ of the set so described: in terms
of the reduced equilibrium distributions~(1), these are obtained by
making the outside potential $V$ very large, whereas in (2) the same
elements play the role of the extreme boundary in the sense of
convex sets.

This suggestive picture loses a bit of its appeal when we consider
more general, less ``extreme'' averages of the Hamiltonian $H$:
rather than evaluating $H$ on a single external configuration $y\in
H$, we can also consider
$$H^\rho(x)= \sum_{y\in\Ybd} \rho(y)\,H(x,y)
\quad, \deqno(ahc)$$
where $\rho$ is a probability distribution on $\Ybd$. Such boundary
conditions arise naturally when considering random systems. They
will also appear when we couple classical systems to quantum outer
systems. One can easily find examples in which the equilibrium
states of such averaged Hamiltonians $H^\rho$ are {\it not}
contained in the convex set described above (see \Exa{C4C3} for an
illustration).

Besides equilibrium distributions, {\it ground states} can be
considered as well. A distribution $\mu$ is a global ground state if
it minimizes the energy functional $\mu\mapsto\mu(\Htot)$. As this
functional is obviously affine, the ground states form a face of the
simplex of probability measures on $X\times Y$, and the same is of
course true for their restriction to the inner system $X$. By an
argument quite similar to that for equilibrium states, one verifies
that the set of reduced ground states, when $Y$ and $V$ vary as
in~\eq(hcc), is still a face of the probability measures on $X$. A
configuration $x_0\in X$ is an extreme reduced ground state iff
there is a $y_0\in\Ybd$ such that
$$H(x,y_0)\geq H(x_0,y_0)\quad \hbox{for all }x\in X\quad.
$$
The situation is quite analogous to that for equilibrium states: in
order to characterize reduced ground states, it suffices to minimize
the energy of the averaged Hamiltonians $H^y$~\eq(ahpc) where $y$
runs over the configurations of the boundary region $\Ybd$.

\bgsection 3. Extension to quantum systems

There are two essential properties of classical systems that
underlie the description of equilibrium states of global systems in
terms of local Gibbs distributions. Firstly, all probability
distributions on a composite system are convex combinations of
product measures, namely the point measures described by pairs of
configurations. This allows conditioning as in equation \eq(dlr).
Secondly, for functions $f$ and $g$, $\exp(f+g)= \exp f\, \exp g$.
Both properties totally fail when we pass to quantum systems.

Quantum systems are described in terms of algebras of observables
rather than configurations. The observables of a discrete, fully
quantum mechanical system form a matrix algebra, say, the algebra
$\M_d$ of complex $d\times d$-matrices. On the other hand, the
observables of a discrete classical system with finite configuration
space $X$ are the complex-valued functions $\C(X)$ on $X$. This is
an Abelian algebra, which can be identified with the diagonal
$d\times d$ matrices, $d$ being the number of points in $X$. In this
paper, we only consider finite-dimensional algebras of observables.
All such algebras decompose into sums of matrix algebras and may
also describe quantum-classical hybrid systems. Probability
distributions are expressed in algebraic language as linear
functionals, called {\it states}, which assign to each observable
its expectation value. On a finite-dimensional algebra $\A$ all such
functionals are of the form
$$ \omega(A)=\tr(D_\omega A)
\quad,$$
where the density matrix $D_\omega$ is positive, and
$\tr D_\omega=1$.

We consider systems split into an ``inner'' and an ``outer''
subsystem. On the level of the algebras, this translates into a
tensor product structure of the global algebra of observables $\A=
\Ain\ot\Aout$: the elements of $\A$ can be written as finite linear
combinations of elementary tensors $A\ot B$, $A\in\A,\ B\in\B$.
$A\mapsto A\ot B$ and $B\mapsto A\ot B$ are linear and the
multiplication in $\A$ is given by
$$(A_1\ot B_1)\, (A_2\ot B_2)= A_1\,A_2\ot B_1\,B_2
\quad.$$
As in the classical case, we will assume that we have a complete
knowledge of the interaction of the particles in the inner system
with the exterior. More precisely
$$\eqalignno{
  H\in\Ain\ot\Abd\quad
  &\hbox{is a fixed Hermitian operator and} \cr
  \Htot= H+ \idty\ot V,\quad
  &V=V^*\in\Aout\supset\Abd
\quad. &\deqal(qqh)} $$
Here, $\Abd$ denotes the smallest $\ast$-subalgebra of $\Aout$ such
that $H\in\Ain\ot\Abd$. For classical systems $\Abd$ is precisely
$\C(\Ybd)$. The equilibrium state $\om$ of the global system is
given by the Gibbs density matrix
$$ D_\om= \Exp{-\Htot}/\Z
\quad. \deqno(gibbsd)$$
The normalization factor $\Z$ is the canonical partition function. We
can restrict the state $\om$ to the inner system to obtain the state
$\omin$ on $\Ain$:
$$\omin(A)= \om(A\ot\idty),\quad A\in\Ain
\quad. \deqno(omr)$$
The density matrix of $\omin$ is the partial trace of the one for
$\om$ with respect to $\Aout$. We now define the set $\SRE$ of
reduced equilibrium states
$$\SRE= \overline{\Set\Big{\omin
              \stt \om\ \hbox{equilibrium state for   }\Htot\
                        \hbox{as in~\eq(qqh)}}}
\quad, \deqno(sre)
$$
where the bar denotes the closure.
The set $\SRE$ is completely determined by $H$ and we will sometimes
explicitly refer to that dependence by writing $\SRE=\SRE(H)$.
By the same argument as in \Sec2, $\SRE$ is a convex set and
we are interested in the following question: can one give a simple
description of $\SRE$ in terms of $H$?

There are two natural ways of averaging the Hamiltonian $H$: either
against pure or against arbitrary states $\rho$ on $\Abd$ (compare with
\eq(ahpc) and \eq(ahc)). In either case, the averaged Hamiltonian is
obtained as
$$\condex\rho(H)= \sum_i \rho(B_i)\,A_i
\quad,$$
where $H=\sum_i A_i\ot B_i$ with $A_i\in\Ain$ and
$B_i\in\Abd\subset\Aout$.
The sets of Gibbs states on $\Ain$ corresponding to these averaged
Hamiltonians will be denoted by
$$\eqalignno{
 \SAHp&= \Set\Big{\om\stt \hbox{equilibrium state for }
  \condex\rho(H)\quad,\ \rho\ \hbox{pure}\quad}
&\deqal(sahp)\cr
\noalign{and}
 \SAH &= \Set\Big{\om\stt \hbox{equilibrium state for }
  \condex\rho(H)\quad,\ \rho\ \hbox{arbitrary}}
\quad.
&\deqal(sah)}$$

In~\cite{Israel}, the set of pure states on $\Abd$ is considered as
a candidate for quantum boundary conditions. As in the classical
case, this choice can be made plausible by considering
perturbations of the Hamiltonian by large potentials: let $P$ be a
minimal projection in $\Aout$ and consider the Hamiltonians
$\Htot(\lambda)= H+ \lambda\,\idty\ot P$. Note that $P$, taken as a
density matrix, also determines a pure state $\rho$ on $\Aout$. We
claim that, as $\lambda\to-\infty$, the restrictions of the
equilibrium states for $\Htot(\lambda)$ to $\Ain$ converge to the
equilibrium state for $\condex\rho(H)$. In order to verify this, we
consider $\lambda^{-1}H$ as a small perturbation of
$-\lambda\idty\otimes P$, obtaining eigenvectors
$\psi_\alpha(\lambda)$, and eigenvalues $\eta_\alpha(\lambda)$, which
depend analytically on $\lambda^{-1}$. Note that the eigenspaces of
$\idty\otimes P$ are highly degenerate, and we must chose the basis
in $0$\th order such that $\psi_\alpha(\infty)$, $\alpha=1,\ldots,d$
is the eigenbasis of the operator
$(\idty\otimes P)H(\idty\otimes P)$ and the remaining
$\psi_\alpha(\infty)$ are an eigenbasis of
$(\idty\otimes(\idty-P))H(\idty\otimes(\idty-P))$. Let
$\eta'_\alpha(\lambda)$ denote the eigenvalues of these operators.
Then the equilibrium state $\omega^\lambda$ is given by
$$ \omega^\lambda(A)
       = {1\over\Z(\lambda)}\sum_\alpha
          e^{-\lambda\eta_\alpha(\lambda)}
          {\langle\psi_\alpha(\lambda),\,
                      A\psi_\alpha(\lambda)\rangle
           \over\norm{\psi_\alpha(\lambda)}^2}
\quad.$$
The exponent is
$$\eqalign{
  \lambda\eta_\alpha(\lambda)
       =\cases{-\lambda+\eta'_\alpha(\lambda)+\Order(\lambda^{-1})
               &for $\alpha=1,\ldots,d$\cr
                 \qquad \eta'_\alpha(\lambda)+\Order(\lambda^{-1})
               &for $\alpha>d$\cr}
\quad.}$$
Hence the terms with $\alpha>d$ are negligible in the sum and,
after cancelling the factor $e^\lambda$ between the sum and
$\Z(\lambda)$, we can take the limit $\lambda\to-\infty$, obtaining
$$\eqalign{
 \omega^\infty(A)
         &={1\over\Z(\infty)}\sum_{\alpha=1}^d
           e^{-\eta'_\alpha(\infty)}
           \langle\psi_\alpha(\infty),\,
                      A\psi_\alpha(\infty)\rangle\cr
   &=(1/\Z) \tr e^{-(\idty\otimes P)H(\idty\otimes P)}A
\quad.}$$
An infinite system generalization of this result can be found
in~\cite{RelEnt}.
Now, if $P$ is the one-dimensional projection onto the vector $\phi$,
the $1$-eigenspace of $(\idty\otimes P)$ consists precisely of the
vectors $\psi\otimes\phi$ where $\psi$ is in the Hilbert space of
the inner system. By identifying $\psi\otimes\phi$ with $\psi$, the
operator $(\idty\otimes P)H(\idty\otimes P)$ becomes
$\condex\rho(H)$, hence $\omega^\infty$ is the Gibbs state with this
Hamiltonian, as claimed.

As any state on $\Abd$ can be extended to a pure state on $\Aout$
for some suitable embedding of $\Abd$ into a $\Aout$, it follows
that
$$\conv\SAH\subset\SRE
\quad.\deqno(SAHinSRE)$$
Obviously, $\SAHp\subset\SAH$ and the states in $\SAHp$ can be
obtained by restricting in the construction of above to the case
$\Aout=\Abd$.

In the classical case we found that $\SRE=\conv\SAHp$. In \Sec6
we will examine whether the converse inclusion of \eq(SAHinSRE)
holds, \ie whether we have
$$\SRE\subset\conv\SAH
\quad.\deqno(DLR)$$
We will refer to this as the {\it DLR-inclusion}. By \eq(SAHinSRE)
it is in fact equivalent to the equality $\SRE=\conv\SAH$.

A state $\om$ of the global system is a {\it ground state} of
$\Htot$ if it minimizes the energy $\om(\Htot)$. For a fixed
$\Htot$, the ground states form a closed convex set, which is even a
face of the state space.  Restricting a pure state of the global
system $\Ain\ot\Aout$ to the inner system will in general destroy
its purity. The set $\SRG$ of reduced ground states is similar to
$\SRE$
$$\SRG= \overline{\Set\Big{\omin
          \stt \om\ \hbox{ground state for }\Htot\
                     \hbox{as in~\eq(qqh)}}}
\quad. \deqno(srg) $$
It is a closed convex set that contains the convex hulls of the sets
$\GAHp$ and $\GAH$ which are defined as the sets of ground states
on $\Ain$ with respect to the averaged Hamiltonians
$\condex\rho(H)$, where $\rho$ varies either over the pure or over
the general states on $\Abd$.

\bgsection 4. Energy-entropy inequalities

An equilibrium state can be characterized as a solution of the
variational principle of thermodynamics:
$$\om\mapsto F(\om)= U(\om)- S(\om)$$
attains its minimum at the Gibbs state. $F$ is the free energy, it
is the difference between the internal energy $U(\om)$ and the
entropy $S(\om)= -\tr D_\om\ln D_\om$. It is possible to
characterize equilibrium states by a set of inequalities that
express in a differential form that the Gibbs state minimizes the
free energy~\cite{EnEnt}. Consider a quantum system described by the
$d\times d$ matrices $\M_d$ and let $H$ be the Hamiltonian of the
system. For any observable $A\in\M_d$, the Gibbs state $\om$
satisfies the ``energy-entropy inequalities''
$$\om(A^*[H,A])\geq \om(A^*A)\, \ln{\om(A^*A)\over\om(A\,A^*)}\quad.
\deqno(eeq)$$
Conversely, if~\eq(eeq) holds for any choice of $A\in\M_d$, then
$\om$ is the Gibbs state defined by $H$. The main advantage of these
inequalities, which express the energy-entropy balance in equilibrium,
is that they involve the Hamiltonian in a linear way.

Consider now again the situation of \Sec3. Any equilibrium state
$\om$ of the global system will satisfy the energy-entropy
inequalities \eq(eeq). In particular, the inequalities must hold for
$A\in\Ain$ and therefore:
$$\om(A^*[\Htot,A])= \om(A^*[H,A])\geq \om(A^*A)\,
  \ln{\om(A^*A)\over\om(A\,A^*)},\quad A\in\Ain
\quad. \deqno(eer)$$
Note that the left hand side involves only the restriction of
$\omega$ to $\Ain\otimes\Abd$, whereas the right hand side involves
only the restriction to $\Ain$. In either case, this condition is
``local'' in the sense that it does not require knowledge of the
global state on $\Aout$. Moreover, if this condition is satisfied
for $\Ain$ varying over the observable algebras of {\it arbitrary}
bounded regions, the inequalities \eq(eeq) hold on the whole
algebra, and hence $\omega$ must be an equilibrium state. Thus at
least some of the key requirements for a useful replacement of the
DLR-equations in the quantum case are satisfied by $\SEE$.

We denote by $\SEE$ the set of states on $\Ain$ defined by
$$\SEE= \Set\Big{\om\stt \exists\,\om\,\tilde{\hbox{}}\
\hbox{extending $\om$ and satisfying~\eq(eer)}}
\quad.$$
The set $\SEE$ is a closed convex set and it contains $\SRE$. Indeed,
in order to show convexity, let $\om$ and $\om'$ be states in
$\SEE$. We can then find extensions $\om\,\tilde{\hbox{}}$ and
$\om'\,\tilde{\hbox{}}$ on $\Ain\ot\Abd$ that satisfy
$$\om^{(\prime)}\,\tilde{\hbox{}}\,(A^*[H,A])\geq
  \om^{(\prime)}(A^*A)\,
  \ln{\om^{(\prime)}(A^*A)\over\om^{(\prime)}(A\,A^*)},\quad
  A\in\Ain
\quad.$$
For $\lambda\in[0,1]$, $\lambda\,\om\,\tilde{\hbox{}}+
(1-\lambda)\,\om'\,\tilde{\hbox{}}$ extends $\lambda\,\om+
(1-\lambda)\,\om'$ and for $A\in\Ain$:
$$\eqalign{
  &\bigl(\lambda\,\om\,\tilde{\hbox{}}+
  (1-\lambda)\,\om'\,\tilde{\hbox{}}\,\bigr)\bigl(A^*\,[H,A]\bigr) \cr
  &\qquad= \lambda\,\om\,\tilde{\hbox{}}\,(A^*\,[H,A])+
  (1-\lambda)\,\om'\,\tilde{\hbox{}}\,(A^*[H,A]) \cr
  &\qquad\geq \lambda\,\om(A^*A)\, \ln {\om(A^*A)\over\om(A\,A^*)}+
  (1-\lambda)\,\om'(A^*A) \ln {\om'(A^*A)\over\om'(A\,A^*)} \cr
  &\qquad= -\Bigl(\lambda\,\om(A^*A) \ln
  {\om(A\,A^*)\over\om(A^*A)}+ (1-\lambda)\,\om'(A^*A) \ln
  {\om'(A\,A^*)\over\om'(A^*A)}\Bigr) \cr
  &\qquad\geq -\bigl(\lambda\,\om(A^*A)+
  (1-\lambda)\,\om'(A^*A)\bigr)\, \ln {\lambda\,\om(A\,A^*)+
  (1-\lambda)\,\om'(A\,A^*)\over \lambda\,\om(A^*A)+
  (1-\lambda)\,\om'(A^*A)} \cr
  &\qquad= \bigl(\lambda\,\om(A^*A)+ (1-\lambda)\,\om'(A^*A)\bigr)\,
  \ln {\lambda\,\om(A^*A)+ (1-\lambda)\,\om'(A^*A)\over
  \lambda\,\om(A\,A^*)+ (1-\lambda)\,\om'(A\,A^*)}
\quad.}$$

There are versions of the energy-entropy inequalities for classical
and hybrid systems too. The inequalities~\eq(eeq) are of no use in
this case because in a classical system all commutators vanish and
$A^*A=AA^*$. Hence \eq(eeq) is trivially satisfied as $0\geq0$.
We therefore need a more general type of inequality that also covers
hybrid systems.
Consider, \eg a coupled classical-quantum system described by the
observable algebra $\C(X)\otimes\A$. If $e_x\in X$ denotes the
function which is $1$ at $x$ and $0$ otherwise, we can expand the
observables $A$ of this system as
$$A= \sum_{x\in X} e_x\otimes A(x)
\quad,\deqno(oh)$$
where the $A(x)$ are elements of $\A$. Thus observables are
identified with $\A$-valued functions on $X$, and the algebraic
operations are defined pointwise in $x$. In particular, $A$ is
Hermitian iff $A(x)^*=A(x)$ for all $x$, and the identity element
is the function which is equal to $\idty\in\A$ for all $x$.
A state $\om$ on such a system is of the form
$$\om= \sum_{x\in X} \delta_x\ot\om^x
\quad,\deqno(sh)$$
where $\delta_x$ is the evaluation at $x$ and the $\om^x$ are
positive functionals on $\A$.
The expectation of an observable~\eq(oh) in the state~\eq(sh) is
computed as
$$\om(A)= \sum_{x\in X} \om^x(A(x))
\quad.$$
With the identity element for $A$, we find the normalization
condition
$$\sum_{x\in X} \om^x(\idty)= 1
\quad.$$

The useful energy-entropy inequalities in this situation are
$$\om^x\Big(A^*\bigl(H(y)\,A- A\,H(x)\bigr)\Big)
     \geq \om^x(A^*A)\, \ln{\om^x(A^*A)\over\om^y(A\,A^*)}
\quad, \deqno(eeh)$$
$x,y\in X$ and $A\in\A$.
{}From this, we find restricted energy-entropy inequalities analogous
to \eq(eer). Assuming that the inner system is the classical
subsystem, \ie $\Ain=\C(X)$, we have to evaluate \eq(eeh) for
$A=\alpha\idty$. Thus $A\,A^*=A^*A$, and a factor
$\abs\alpha^2$ can be cancelled from \eq(eeh). Hence, for a
classical inner system, the suitable definition for $\SEE$ is the
set of all probability measures $\mu$ on $X$ for which there exist
non-negative functionals $\om^x$ on $\A$ with $\mu(x)= \om^x(\idty)$
and such that
$$\om^x\big(H(y)- H(x)\big)\geq \mu(x)\,
  \ln{\mu(x)\over\mu(y)},\quad x,\ y\in X
\quad. \deqno(EEhybrid)$$

\bgsection 5. Systems with a classical interior algebra

With the exception of \Exa{Q2Q2}, all examples in \Sec6 will be ``hybrid
systems'' with a classical inside system with $n=$ finitely many
configurations, \ie we have $\Ain=\Cc n$. The interaction is given
by a Hermitian element $H\in\Ain\otimes\Aout$, \ie  according to
\eq(oh), by a Hermitian-valued function $x\mapsto H(x)\in\Aout$.
The subalgebra $\Abd\subset\Aout$ is the algebra generated by
$\idty$, and all $H(x)$. The perturbation $\idty\otimes V$ by a
potential $V=V^*\in\Aout$ is represented by the corresponding
$x$-independent function. The reduced equilibrium states of the
total Hamiltonian $\Htot=H+\oneten V$ are then the probability
distributions
$$ \mu\RE^{V}(x)
    = {1\over \Z} \tr\Bigl(\Exp{-(H(x)+V)}\Bigr)
\qquad,\quad x=1,\ldots,n
\quad,\deqno(exeqRE)$$
where $\Z$ is the normalization factor making this a probability
distribution. If $\rho$ is a state on $\Aout$, the averaged
Hamiltonian $\condex\rho(H)$ is the function $h=\condex\rho(H)\in\Cc
n$ taking the value $h(x)=\rho(H(x))$ at $x$. (Note that only the
restriction $\rho\rstr\Abd$ enters this expression). This leads to
a probability distribution of the form
$$ \mu\AH^\rho(x)={1\over \Z}\ \Exp{-\rho(H(x))}
\qquad,\quad x=1,\ldots,n
\quad.\deqno(exeqAH)$$
General states $\omega$ on $\Ain\otimes\Aout$ are given by (not
normalized) positive functionals $\omega^x$ on $\Aout$, as in
\eq(sh). By \eq(EEhybrid), the
energy-entropy inequalities for such a state mean that
$$ \omega^x\Bigl(H(y)-H(x)\Bigr)
  \geq \mu\EE(x)\,\ln{\mu\EE(x)\over\mu\EE(y)}
\qquad,\quad x,y=1,\ldots,n
\quad.\deqno(exeqEE)$$
Of course, the restricted state $\mu\EE\in\SEE$ is related to the
functionals $\om^x$ by
$\mu\EE(x)= \omega^x(\idty)$, with $\omega$
satisfying this inequality.

It is sometimes convenient to work with the logarithms of
probabilities, \ie to describe a state $\mu$ on $\Ain$ by the
$n-1$ numbers
$$ (\lntw\mu)(x)
      =\ln\bigl(\mu(x)\bigr)-\ln\bigl(\mu(1)\bigr)
\quad,\quad\hbox{for}\quad x=2,\ldots,n
\quad.\deqno(lntw)$$
The image of the various sets ${\cal E}_{\rm XX}\subset\Rl^{n-1}$
under this map will be denoted by  $\lntw{\cal E}_{\rm XX}$. In this
logarithmic scale $\SAH$ becomes
$$ \lntw\SAH=\Set\Big{\xi\in\Rl^{n-1}
             \stt \xi_x=\rho\bigl(H(1)-H(x)\bigr) }
\quad,\deqno(lnSAH)$$
where $\rho$ runs over all states of $\Abd$. This is an affine
image of the state space of $\Abd$. Note that this set is always
convex but, due to the non-linearity of $\exp$, $\SAH$ itself is not
(compare \Fig{C4C3}).

When $\Ain$ is classical, we can also give a fairly explicit
description of $\SEE$. Since we want to interpret
\eq(exeqEE) as a condition for the probability distribution
$\mu\EE(x)=\omega^x(\idty)$, it is convenient to write the
positive linear functionals $\omega^x$ determining $\omega$ as
$\omega^x=\mu\EE(x)\cdot\rho^x$, where the $\rho^x$ are now
states on $\Aout$. Note that no $\mu\EE(y)$ can vanish: since
$\omega\neq0$, some $\mu\EE(x)$ must be non-zero, and by
\eq(exeqEE) $\ln(\mu\EE(x)/\mu\EE(y))$ is not infinite. Hence
we may divide \eq(exeqEE) by $\mu\EE(x)$ and we find
$$    \ln\mu\EE(y) +\rho^x(H(y))
   \geq\ln\mu\EE(x) +\rho^x(H(x))
\quad.\deqno(lncone)$$
In these terms, the energy-entropy condition on $\mu\EE$ is that,
for every $x$, there is a state $\rho^x$ on $\Aout$ such that
\eq(lncone) holds for all $y$. We consider first the case $x=1$.
Then, using the equations \eq(lntw) and \eq(lnSAH), we conclude
that there is a vector $\xi\in\lntw\SAH$,
namely $\xi_y=\rho^x(H(1)-H(y))$, such that $\lntw\mu\EE-\xi$ lies
in the cone $C_1$ of coordinatewise positive vectors in $\Rl^{n-1}$.
In shorthand notation, this is written as
$\lntw\mu\EE\in\lntw\SAH+C_1$, where, as usual, the sum on the right
is to be interpreted as
$\set{\xi+\eta\stt\xi\in\lntw\SAH,\ \eta\in C_1}$. The conditions
for $x>1$ can be written in a similar form after subtracting
$\ln\mu\EE(1) +\rho^x(H(1))$ from both sides of the inequality
when $x\neq y$. Thus, with the cones
$$ C_x=\cases{ \set{\xi\in\Rl^{n-1}\stt \forall y\ \xi_y\geq0}
                  &$x=1$\cr
\noalign{\vskip4pt}
               \set{\xi\in\Rl^{n-1}
                 \stt \forall y\ \xi_y\geq\xi_x
                       \ \hbox{and}\ \xi_x\leq0  }
                  &$x>1$}
\quad,\deqno(SEEcones)$$
we get the formula
$$ \lntw\SEE=\bigcap_{x=1}^n(\lntw\SAH+C_x)
\quad.\deqno(hybSEE)$$
This construction is illustrated below in \Exa{C3Q2} (see
\Fig{C3Q2b}). An amazing feature of this set is that both
$\lntw\SEE$ and its exponentiated version $\SEE$ are convex (see
\Sec4 for the second statement). This seems to be in conflict
with the non-linearity of the map $\lntw$. However, this conflict is
resolved by considering the special form of the cones \eq(SEEcones):
the cone $C_x$ corresponds to that subset of the state space in
which $\ln\mu(x)\leq\ln\mu(y)$ for all $y$. Such inequalities
survive exponentiation, \ie the cone $C_x +\lntw\mu$ at $\mu$
corresponds in the state space to the set of probability
distributions $\mu'$ such that $\mu'(x)/\mu(x)\leq\mu'(y)/\mu(y)$
for all $y$, which is a convex set.

\bgsection 6. Examples

Since our main aim is to demonstrate the failure of the
DLR-inclusion in the quantum case, our main results are in the form
of examples demonstrating the success or failure of these ideas in
various situations. We have chosen a succession of examples to
illustrate various degrees of failure of the DLR-inclusion
$\SRE\subset\conv\SAH$. Mathematically, the strongest result is the last
example, in which both the inside system $\Ain$, and the boundary
algebra $\Abd$ are classical, and $\Aout$ is the algebra of
$3\times3$-matrices. However, this example is rather indirect, so we
will give simpler, direct examples as well, hoping to help the
reader to develop an intuition of what exactly goes wrong. We
emphasize that, while we treat only finite systems, our results are
equally valid for infinite outside systems. In fact, every example
that we give can easily be enlarged to an example with infinite
outside.  In the same spirit we are looking mostly for examples with
classical $\Ain$: usually these can be expanded easily to examples
with non-commutative $\Ain$. Moreover, they are simpler to treat
and, being closer to the classical DLR-situation, they bring the
failure of the quantum generalization into sharper focus. In
producing the examples we found the help of a computer algebra
program~\cite{Mathematica} a very valuable tool.

The symbols in the heading of each example are a shorthand for the
type of algebras used. Thus ``Q5C3'' is an example with a quantum
inside with $\Ain=\M_5$, and an interaction $H$ contained in a
subalgebra of $\Ain\otimes\Abd$, which is isomorphic to
$\Ain\otimes\Cc3$, \ie the inside system interacts with $3$
classical configurations. Of course, the whole algebra $\Aout$ is
never assumed to be classical, since then the DLR-theory would
simply be valid. If the number of configurations or Hilbert space
dimensions is irrelevant, we just write ``n''. Each example begins
with a short description of the model and of the claims about various
inclusions to be seen in this example (set in {\it italics}),
followed by the verification of these claims and some additional
remarks.

\EXA C4C3:C4C3; $\Aout$ {\rm classical}:
With this example, where $\Aout$ is Abelian, we demonstrate that
$\conv\SAH\not\subset\conv\SAHp$. It also serves to illustrate the
lack of any definite convexity properties of the map which takes
Hamiltonians to Gibbs distributions (see \Fig{C4C3}).
Explicitly, the interaction is given by the matrix
$H(x,y),\  x=1,2,3,4,\quad y=1,2,3$, as
$$ H=\pmatrix{0&0&0\cr  1&-1&0\cr
                  0&1&-1\cr  1&0&-1}
\quad.$$

The numerical evaluation in this example is straightforward, and the
result is shown in \Fig{C4C3}. We include this example mainly to
illustrate a remark made in \Sec2, namely that, due to the
non-linearity of the exponentiation map, general averaged
Hamiltonians \eq(ahc) do not give equilibrium states inside the
DLR-triangle $\SRE$, which is also marked in the figure.

\EXA C2Qn:C2Qn:
When the inside system is a single Ising spin and the outside is
arbitrary, the DLR-inclusion holds, namely
$$ \SRE=\SAH=\SAHp
\quad.$$

The Hamiltonian is specified by two Hermitian $H(1),H(2)\in\M_n$.
Then, by \eq(lnSAH), $\lntw\SAH$ is just the set of numbers
$\rho(H(1)-H(2))$, where $\rho$ ranges over all states of $\Aout$.
Thus $\lntw\SAH$ is the interval $\bracks{\eta_-,\eta_+}$, with
endpoints given by the largest and smallest eigenvalues of
$H(1)-H(2)$,
\ie the smallest $\eta_+$ and the largest $\eta_-$ such that
$\eta_-\idty\leq H(1)-H(2)\leq \eta_+\idty$. Moreover, the same
interval is obtained if we restrict the states $\rho$ to be pure,
\ie $\SAH=\SAHp=\bracks{\eta_-,\eta_+}$.

On the logarithmic scale,
$\mu\RE^V\in\SRE$ from~\eq(sre), is given by the single number
$$    \eta(V)= \ln\trexp-(H(2)+V)-\ln\trexp-(H(1)+V)
\quad.$$
Now the function $A\mapsto\ln\trexp(-A)$ is monotonely decreasing
with respect to the operator ordering for $A$. Moreover,
$\ln\trexp(-A-\eta\idty)=\ln\trexp(-A)-\eta$.
Since $H(2)\leq H(1)-\eta_-\idty$, we get $\eta(V)\geq\eta_-$.
Similarly, we get the upper bound $\eta(V)\leq\eta_+$. Hence
$\SRE\subset\bracks{\eta_-,\eta_+}=\SAHp$, and, from the general
arguments in \Sec3 concerning large potentials, we get the inclusion
$\SAHp\subset\SRE$, and hence equality.

\EXA Q2Q2:Q2Q2:
We now replace the Ising spin of the inside system by a quantum
mechanical spin-${1\over2}$, \ie we take $\Ain=\M_2$. For the
outside, we will also have $\Aout=\M_2$, and define the Hamiltonian
by
$$ H(\alpha,\gamma)
      = \sigma_1\otimes\sigma_1+\sigma_2\otimes\sigma_2
        +\gamma\,\sigma_3\otimes\sigma_3
        -\alpha\,\sigma_3\otimes\idty
\quad,\deqno(q2q2.H)$$
where $\alpha>0$ and $\gamma\geq0$ are parameters, and $\sigma_i$,
$i=1,2,3$ are the usual Pauli matrices. We claim that, in this
model, the reduced equilibrium state $\omin^0\in\SRE\bigl(\beta
H(\alpha,\gamma)\bigr)$, where the superscript 0 denotes that the
potential $V$ is vanishing, is not contained in
$\conv\SAH\bigl(\beta H(\alpha,\gamma)\bigr)$. Hence the
DLR-inclusion fails in the following cases:
\vskip0pt\nobreak
\vtop{\parskip=0pt\parindent=60pt
\item{(1)} $\beta=3,\ \alpha=.3,\ \gamma=.05$ (see \Fig{Q2Q2})
\item{(2)} $\gamma=0$, and all $\beta>0$,
\item{(3)} for $0\leq\gamma\leq.8$, and  $\beta=\infty$, \ie for
           ground states. }

What makes this operator tractable is that it commutes with
SU(2)-rotations around the $3$-axis. In the joint eigenbasis of
$\sigma_3\otimes\idty$ and $\idty\otimes\sigma_3$ it corresponds to
the matrix
$$H(\alpha,\gamma)
         =\pmatrix{-\alpha+\gamma&0&0&0 \cr
                    0            & -\alpha-\gamma &2&0\cr
                    0            &2&\alpha-\gamma &0\cr
                    0&0&0                         &\alpha+\gamma }
\quad.$$
The partition function for $H(\alpha,\gamma)$ and the expectation
of $\sigma_3\otimes\idty$ are readily computed:
$$\eqalignno{
  \Z(\beta,\alpha,\gamma)
       &= \tr \Exp{-\beta H(\alpha,\gamma)}
        =  2 e^{-\beta\gamma}\ \cosh(\beta\alpha)
         + 2 e^{\beta\gamma}\  \cosh(\beta R_1) \cr
 \omin^0(\sigma_3)
      &={1\over\beta}\, {\partial\over\partial\alpha}
        \ln\,\Z(\beta,\alpha,\gamma)   \cr
      &={{\alpha {e^{\beta (\alpha+2\gamma) }}
              \left( {e^{2\beta R_1}}-1 \right)  +
             R_1{e^{\beta R_1}}\left(e^{2\alpha\beta}-1 \right)}
          \over
          {R_1\left( {e^{\beta(\alpha+2\gamma) }} +
               {e^{\beta R_1}} + {e^{\beta(2\alpha + R_1)}} +
               {e^{\beta(\alpha + 2\gamma  + 2 R_1)}}\right) }}
\quad,&\deqal(q2q2.mRE)}$$
where $R_1=\sqrt{4+\alpha^2}$. Since $H(\alpha,\gamma)$ commutes
with rotations around the $3$-axis, so does $\omin^0$, and hence
it is completely determined by the expectation of $\sigma_3$.

If $\rho={1\over2}\smallpmatrix{1+x_3&x_1-ix_2\cr
                                x_1+ix_2&1-x_3}$, with
$x_i\in\Rl$ and $\sum_ix_i^2\leq1$, is a $2\times2$-density matrix,
we find
$$ \condex\rho(H(\alpha,\gamma))
     =\pmatrix{ -\alpha+\gamma x_3&x_1-ix_2\cr
               x_1+ix_2&  \alpha-\gamma x_3 }
\quad.$$
The equilibrium state $\om\AH^\rho$ of
$\condex\rho(H(\alpha,\gamma))$ can be evaluated with the formula
$$ {e^{-\beta H}\over\tr e^{-\beta H}}
    ={1\over2}\idty- {\tanh\beta\sqrt{\det H}\over2\sqrt{\det H} }\ H
\deqno(2by2tanh)$$
for $H$ a traceless $2\times2$-matrix.
$\SAH$ is now a body in the state space of $\M_2$, parametrized by
the triple $(x_1,x_2,x_3)$ determining $\rho$. It is the
image of an ellipsoid in the space of $2\times2$ Hamiltonians under
the non-linear map \eq(2by2tanh). This map preserves the symmetry
under rotations around the $3$-axis, hence $\SAH$ will also have
this symmetry. \Fig{Q2Q2} shows a section of $\SAH$ in the
$(1,3)$-plane of the state space. It has to be completed by rotating
it around the vertical axis, thus making $\SAH$ the shape of a
mushroom. The point marked on the axis is $\omin^0$. It is
clearly not contained in the convex hull of $\SRE$.

At high temperature (small $\beta$) one can expand all
exponentials, and one finds that, in first order in $\beta$ and as
long as $\gamma\neq0$, the convex hull of $\SAH$ does contain
$\omin^0$. However, if $\gamma=0$, the first orders coincide,
and we get
$$ \omega\AH^\rho(\sigma_3)-\omin^0(\sigma_3)
       \geq {\alpha\over3}(2-x_1^2-x_2^2)\ \beta^3
          +\Order(\beta^5)
       >0
\quad.$$
Hence the DLR-inclusion fails even for small $\beta$, and it can
be seen numerically that, with $\gamma=0$, it fails for all
$\beta\geq0$.

In the opposite direction, as $\beta$ grows, the points in $\SAH$
move radially out to the extreme boundary of the state space and, in
the limit, we obtain the pure ground states of the Hamiltonians
$\condex\rho(H(\alpha,\gamma))$.  Hence, for $\beta=\infty$
$$ \omega\AH^\rho(\sigma_3)
     = {\alpha-\gamma x_3\over
        \sqrt{x_1^2+x_2^2+(\alpha-\gamma x_3)^2}}
      \geq{{\sqrt{ \alpha^2 - \gamma^2}}\over
           {\sqrt{1 + \alpha^2-\gamma^2}}}
\quad,$$
where the second expression is the minimum of the first with respect
to variation of the $x_i$, attained for $x_1^2+x_2^2=1-x_3^2$ and
$x_3=\gamma/\alpha$. The limit $\beta\to\infty$ of $\omin^0$ can
be obtained directly from \eq(q2q2.mRE): of all the exponents
appearing in this equation, $\beta(\alpha+2\gamma+2R_1)$ is clearly
the largest, so the limit is $\alpha/R_1$, or
$$ \omin^0(\sigma_3) = {\alpha\over\sqrt{4+\alpha^2}}
\quad.$$
Hence, for all $\rho$, we have
$$\omega\AH^\rho(\sigma_3) > \omin^0(\sigma_3)
\quad\hbox{for $\beta=\infty$
        and $0\leq\gamma<{\sqrt3\over2}\alpha$}\quad.$$
Thus the DLR-inclusion fails in general for quantum ground states.
For this conclusion it is crucial that both systems are
non-classical. In our further examples we will always take $\Ain$ Abelian.
The pure states of the tensor product $\Ain\otimes\Aout$, such as
the extremal ground states of $H$, are then product states and can
be found by minimizing $\condex\rho(H)$ for each $\rho$. Thus, for
classical $\Ain$, the ground state DLR-inclusion holds.

\EXA QCQ:QnCm; $\Aout= \Abd\otimes\B$:
The DLR-inclusion $\SRE= \conv\SAHp$ holds when the boundary algebra
is purely classical and when the exterior algebra $\Aout$ factorizes
into a tensor product $\C(\Ybd)\ot\B$, $\B$ arbitrary.

{}From the general argument of \Sec2 we always have
$\conv\SAHp\subset\SRE$. Conversely, with $\Htot= H+ \oneten V$, $H$
and $V$ functions from $\Ybd$ to $\Ain$ and $\B$, we have for any
$A\in\Ain$
$$\sum_{y\in\Ybd} \tr_{\Ain\ot\B} \left(\Exp{-(H+V)}\,A\right)=
  \sum_{y\in\Ybd} \tr_{\B} \left(\Exp{-V(y)}\right) \tr_{\Ain}
  \left(\Exp{-H(y)}\,A\right)
\ .$$
But this precisely means that $\SRE\subset\conv\SAHp$. From
\Exa{C4C3} it becomes clear that generally $\conv\SAHp$ is strictly
contained in $\conv\SAH$. Therefore the inclusion
$\conv\SAH\subset\SRE$, that holds if we allow for general
embeddings of $\C(\Ybd)$ in an exterior algebra $\Aout$, is reversed
if we restrict ourselves to product systems $\Ain\ot\C(\Ybd)\ot\B$.

A special case of this example is the case where all of $\Aout$ is
classical. The representation of equilibrium states by conditioning
with respect to a classical outside is reminiscent of the work
\cite{Freuden}. In that project one also considers quantum systems
split into inner and outer region. The aim is then to represent
general quantum states as integrals of states conditioned with
respect to a classical subalgebra in the outside region. In the
present, finite-dimensional setting, this is a trivial operation.

\EXA C3Q2:C3Q2; $\Aout= \M_2$:
By \Exa{C2Qn}, the smallest non-trivial case of a classical inside
algebra is $\Ain=\Cc3$ and we stay with $\Aout=\M_2$ \ie we
consider only the subset $\SRE(\M_2)\subset\SRE$ of reduced
equilibrium states coming from potentials $V\in\M_2$. Thus $H$ is
given by three $2\times2$-matrices, which we choose as
$$ H(1)=0       \quad,\quad H(2)={3\over2}\sigma_3
\quad,\quad\hbox{and}\quad H(3)={3\over2}\sigma_1
\quad.$$
In this example, we will determine the three sets
$\SAH\subset\SRE(\M_2)\subset\SEE$ explicitly and show that both
inclusions are strict (see \Fig{C3Q2a}).

The averaged Hamiltonians $\condex\rho(H)$ depend only on the two
expectations $\rho(\sigma_3)$ and $\rho(\sigma_1)$, and therefore
form a two-dimensional disc in the space of Hamiltonians. Note that,
because $\rho(\sigma_2)$ does not enter, we get the same disc
when Hamiltonians are averaged only by pure states $\rho$. The disc of
Hamiltonians is (up to a factor) identical with $\lntw\SAH$  (see
the circle in \Fig{C3Q2a}).

In the logarithmic scale $\mu\RE^V$ is given by a point in the
plane with coordinates
$$ \eta(V)= \lntw\mu\RE^V
          =\pmatrix{\ln\trexp(-H(2)-V)-\ln\trexp(-V) \cr
                    \ln\trexp(-H(3)-V)-\ln\trexp(-V) \cr}
\quad.$$
By choosing some random $V$, it is easy to convince oneself that
in this example $\SRE(\M_2)\not\subset\SAH$, \ie we get points outside the
circle in \Fig{C3Q2a}.
The precise determination of $\SRE(\M_2)$ is more difficult. Consider the
map $V\mapsto\eta(V)$. We know that,
as $\lambda\to\infty$, $\eta(\lambda V)$
approaches a point in $\SAH$. Therefore the boundary points of
$\SRE(\M_2)$ must be of the form $\eta(V)$ with finite $V$,
and at such points the map $\eta$  must be singular, more precisely,
the rank of the Jacobian must be one. It turns out that this can
happen only in the plane $V=x\sigma_1+z\sigma_3$, along three
disjoint arcs. Each of these arcs extends to infinity, and describes
a piece of the boundary of $\SRE(\M_2)$ connecting with the boundary of
$\SAH$. Solving for the critical points in the $(x,z)$-plane can
only be done numerically and the result is shown in \Fig{C3Q2a}.

In this example we can also evaluate the energy-entropy inequalities
\eq(exeqEE). Using the general method outlined in \Sec5, we
obtain $\lntw\SEE$ as the intersection of the three convex sets shown
in \Fig{C3Q2b}. The result is also included in \Fig{C3Q2a} for
easier comparison.

\EXA C3C2:C3C2:
In the previous example the algebra $\Abd$ was non-Abelian.
We now move still closer to the classical DLR-situation by requiring
this algebra to be Abelian too. We will retain the choice
$\Ain=\Cc3$ of the previous example, but set $\Abd=\Cc2$.
More generally, the considerations of this example are valid if
$\Ain=\Cc3$ and if the three operators $H(x)\in\Aout$, determining the
interaction, are of the form
$$ H(x)=h_0+ a_x h_1+ b_x\idty
\quad,$$
for some Hermitian $h_0,h_1\in\Aout$ and real constants $a_x,b_x$.
We then find that the DLR-inclusion holds, \ie $\SRE=\conv\SAH$.
Moreover,  the inclusion $\SRE\subset\SEE$ is strict (see \Fig{C3C2}).

Without loss of generality, we can take $a_2\geq a_1\geq a_3$.
Then we can find a $\lambda$, $0\leq\lambda\leq1$, such that
$$ H(1)=\lambda H(2) +(1-\lambda) H(3) + \widetilde y \idty
\quad,\deqno(c3c2:h)$$
where $\widetilde y=b_1-\lambda b_2-(1-\lambda)b_3$. Hence, we have
$$ \ln\mu(1)
    =\lambda\, \ln \mu(2) +(1-\lambda)\, \ln \mu(3) - \widetilde y
\quad,\quad\hbox{for }\quad\mu\in\SAH
\quad.\deqno(c3c2:lnAH)$$
Thus $\SAH$ lies on a line in the state space, which appears
straight on the logarithmic scale. The set $\SAH$ is parametrized by the
expectation $\rho(h_1)$ in arbitrary states $\rho$ of $\Aout$ and
the extremal eigenvalues $\eta_{\pm}$ of $h_1$ (\ie the best
constants with $\eta_-\idty\leq h_1\leq\eta_+\idty$) determine the
$\SAH$ as a segment of that line. An alternative description of the
endpoints is by the equation of the line passing through them, or
equivalently, by the coefficients $c_2,c_3$ of the linear inequality
$$ \mu(1)\geq c_2\mu(2)+c_3\mu(3)
\quad,\deqno(c3c2:AH)$$
which becomes an equality precisely at these endpoints. Inserting
the known endpoints $\mu_\pm$ and solving the linear system, we
find
$$\eqalign{
c_2&= {\textstyle
        e^{\eta_- (a_2-a_1) + \eta_+ (a_2-a_1) + (b_2-b_1)}
         \bigl(e^{\eta_- (a_3-a_1)} - e^{\eta_+ (a_3-a_1)}\bigr)
      \over \textstyle e^{\eta_+ (a_2-a_1) + \eta_- (a_3-a_1)}
                     - e^{\eta_- (a_2-a_1) + \eta_+ (a_3-a_1)}}\cr
c_3&={\textstyle
       e^{\eta_- (a_3-a_1) + \eta_+ (a_3-a_1) + (b_3-b_1)}
         \bigl( e^{\eta_+ (a_2-a_1)} -e^{\eta_- (a_2-a_1)}\bigr)
      \over\textstyle  e^{\eta_+ (a_2-a_1) + \eta_- (a_3-a_1)}
                     - e^{\eta_- (a_2-a_1) + \eta_+ (a_3-a_1)}}
\quad.}$$
The only information we will need from these formulas is that,
since $\eta_+\geq\eta_-$ and $a_2\geq a_1\geq a_3$, both constants
are positive. This implies that, for all $\eta$ with
$\eta_-\leq\eta\leq\eta_+$,
$$ e^{-a_1\eta-b_1}
        \geq c_2 e^{-a_2\eta-b_2} +c_3 e^{-a_3\eta-b_3}
\quad.$$
Indeed, this is an equality for $\eta=\eta_+$ and $\eta=\eta_-$ and
follows for the intermediate values by multiplying both sides with
$e^{a_1\eta}$ and invoking the convexity of the exponential
function. Since $\rho(h_1)\in\bracks{\eta_-,\eta_+}$ for all states
$\rho$ on $\Aout$, we find that $\SAH$ is the set of probability
distributions satisfying equation \eq(c3c2:lnAH) and
inequality~\eq(c3c2:AH). (See the thick line in \Fig{C3C2}).

On the other hand, since the interval $\bracks{\eta_-,\eta_+}$
contains the spectrum of $h_1$ and both $(H(1)-H(2))$ and
$(H(1)-H(3))$ are functions of $h_1$, we obtain from \eq(c3c2:h) the
operator inequality
$$  \idty\geq c_2\, e^{H(1)-H(2)}+ c_3\, e^{H(1)-H(3)}
\quad.$$
Combining this with the Golden-Thompson inequality~\cite{GoldThom}, \ie
$\tr\exp(A+B)\leq\tr(\exp(A)\,\exp(B))$, we get
$$\eqalign{
   c_2\tr e^{-H(2)-V}&+c_3\tr e^{-H(3)-V}   \cr
     &= c_2\tr e^{(H(1)-H(2))-(H(1)+V)}
       +c_3\tr e^{(H(1)-H(3))-(H(1)+V)}\cr
     &\leq\tr\left(c_2\, e^{H(1)-H(2)}+ c_3\, e^{H(1)-H(3)} \right)
               e^{-H(1)-V} \cr
     &\leq \tr e^{-H(1)-V}
\quad.\cr}$$
Hence, all probability distributions $\mu\in\SRE$ also satisfy
inequality \eq(c3c2:AH). On the other hand, we may use the convexity
of the function $A\mapsto\ln\trexp(-A)$, together with \eq(c3c2:h)
to obtain that, for $\mu\in\SRE$,
$$ \ln\mu(1)
    \leq\lambda\, \ln \mu(2) +(1-\lambda)\, \ln \mu(3) - \widetilde y
\quad.$$
Thus we have shown the DLR-inclusion $\SRE\subset\SAH$. The
determination of $\SEE$ uses the method described in \Sec5 and
illustrated in \Fig{C3Q2b}.

\EXA CnC3:CnC3:
The previous example might have nourished the hope that, at least for
purely classical interactions ($\Abd$ Abelian), the DLR-inclusion
might survive. What we will show now, however, is that
$\SRE\not\subset\conv\SAH$ in this case. The following Proposition
reduces our question to a well-known failed conjecture, which was
disproved by Gaudin \cite{Gaudi}.

\proclaim Proposition.
Let $\Aout=\M_3$, and denote by $\D_3$ the subalgebra of diagonal
matrices. Suppose that, for every $n$ and every Hamiltonian
$H\in\Cc n\otimes\D_3$, the DLR-inclusion $\SRE\subset\conv\SAH$
holds.
Then, for any Hermitian $3\times3$-matrix $V$, there is a positive
measure $\mu_V$ on $\Rl^3$ such that
$$ \trexp\left(-\smallpmatrix{x_1&0&0\cr0&x_2&0\cr0&0&x_3}
                -V\right)
       =\int\!\!\mu_V(d\lambda_1\,d\lambda_2\,d\lambda_3)
              \ \Exp{-\sum \lambda_i x_i}
\quad.\deqno(Gaudi)$$

\proof:
Let us fix some Hermitian $3\times3$-matrix $V$.
By $E_i\in\D_3$ we denote the matrix with ``$1$'' in the $i$\th place
of the diagonal, and all other entries zero. For $x\in\Rl^3$ we
set $H(x)=\sum_{i=1}^3x_iE_i$.
Consider a finite subset $X\subset\Rl^3$. Then $H(x)$ defines an
interaction $H\in\C(X)\otimes\D_3$. The left hand side of
equation \eq(Gaudi) is $\mu\RE^V(x)$, for $x\in X$. On the other
hand, the average of $H$ with respect to a state on $\M_3$ (or,
equivalently, a state on $\D_3$) is characterized by three numbers
$\lambda_i\geq0$ with $\sum_i\lambda_i=1$, and
$\condex\lambda(H)(x)=\sum_i\lambda_ix_i$. By the inclusion
$\SRE\subset\SAH$ we can find a measure $\mu_V^X$ with finite
support, such that equation \eq(Gaudi) holds for all $x\in X$
with $\mu_V^X$ for
$\mu_V$. Now consider the net of finite subsets
of $X\subset\Rl^3$, ordered by inclusion. We may assume that each $X$
contains the origin, so that each one of the associated measures
$\mu_V^X$ is normalized to the same constant $\trexp(-V)$. Then, by
weak*-compactness, we find an accumulation point $\mu_V$ of this net of
measures and it is clear that equation \eq(Gaudi) holds for this
measure and all $x\in\Rl$.
\QED

We may paraphrase this by saying that, given the validity of the
inclusion $\SRE\subset\SAH$, the expression $\trexp A$, considered
as a function of the diagonal matrix elements of $A$, is the Laplace
transform of a positive measure. This is precisely the statement
investigated by Gaudin \cite{Gaudi} and proven to be wrong by
analyzing the measure $\mu_V$, which is, of course, uniquely
determined by $V$ due to uniqueness of Laplace transforms.

Rather than considering the expression $\trexp A$ as a function of
its $n$ diagonal elements, one can also study this expression along a
single straight line in the space of Hermitian $n\times n$-matrices.
The following conjecture about this situation was formulated by
Bessis, Moussa, and Villani in 1975 \cite{BMV}:

\proclaim Conjecture.
Let $A$ and $B$ be Hermitian $n\times n$-matrices. Then there is a
positive measure $\mu$ such that
$$ \tr \Exp{A+tB}=\int\!\mu(d\beta)\ \Exp{\beta t}
\quad.$$

A, possibly signed, measure $\mu$ satisfying this equation always
exists and is uniquely determined by $A$ and $B$. Hence, the issue is
only the positivity of this measure.
Despite many attempts, this conjecture is still open, even for $n=3$.
It is true for $n=2$ and whenever $A$ has positive matrix
off-diagonal elements in an eigenbasis of $B$ \cite{BMV}.
Moreover, it is true for sufficiently small $A$ \cite{TXP}.
The measure $\mu$ is known to have support in the convex hull of the
eigenvalues of $B$ and to have a positive atomic part, supported by
the eigenvalues themselves. A review of the existing partial results,
together with some new ones, is in preparation \cite{TXP}.

\let\REF\doref
\Acknow
The distinction between the sets $\SAH,\ \SAH$, and $\SEE$
was discussed earlier in an internal report by P.~Vets~\cite{Vets}.
We have benefitted from discussions with Bruno Nachtergaele,
Aernout van~Enter, and Christian Maes. Most of this work was done during
visits of M.F.\ to Osnabr\"uck. He would like to take this
opportunity to express his thanks for the warm hospitality and most
enjoyable cooperation. R.F.W.\ was partly supported by a scholarship
from the DFG (Bonn).

\REF Ara GoldThom  \Jref
     H. Araki
     "Golden-Thompson and Peierls-Bogoliubov inequalities for a
     general von~Neumann algebra"
      Commun.Math.Phys. @34(1973) 167--178

\REF AI GibbsCond  \Jref
     H. Araki, P.D.F. Ion
     "On the equivalence of KMS and Gibbs conditions for states of quantum
     lattice systems"
      Commun.Math.Phys. @35(1974) 1--12

\REF BMV BMV \Jref
    D. Bessis, P. Moussa, M. Villani
    "Monotonic converging variational approximations to the
    functional integrals in quantum statistical mechanics"
    J.Math.Phys. @16(1975) 2318--2325

\REF BR BraRo      \Bref
    O. Bratteli, D.W. Robinson
     "Operator algebras and quantum statistical mechanics"
     2 volumes, Springer Verlag, Berlin, Heidelberg, New York
     1979 and 1981

\REF DS  Dobrushin  \Gref
     R.L. Dobrushin, S.B. Shlosman
     "Constructive criteria for the uniqueness of Gibbs fields"
    \inPr J. Fritz, A. Jaffe, D. Szasz
    "Statistical physics and dynamical systems"
    Birkh\"auser, Boston 1985

\REF FNW FCS \Jref
        M. Fannes, B. Nachtergaele, R.F. Werner
        "Finitely correlated states on quantum spin chains"
        Commun.Math.Phys. @144(1992) 443--490

\REF FV EnEnt \Jref
      M. Fannes, A. Verbeure
      "Correlation inequalities and equilibrium states"
      Commun.Math.Phys. @55(1977) 125--131

\REF FW TXP \Gref
    M. Fannes, R.F. Werner
    ""
    In preparation

\REF FF Freuden \Jref
    K.-H. Fichtner, W. Freudenberg
    "Characterization of states of infinite Boson systems"
    Commun.Math.Phys. @137(1991) 315--357

\REF Gau Gaudi \Jref
    M. Gaudin
    "Sur la transform\'ee de Laplace de ${\rm tr}\, e^{-A}$
    consid\'er\'ee comme fonction de la diagonale de $A$"
    Ann.Inst.Henri Poincar\'e A @28(1978) 431--442

\REF Ger Gerisch \Jref
    T. Gerisch
    "Local perturbations and limiting Gibbs states of quantum
    lattice mean-field systems"
    Helv.Phys.Acta @67(1994)585--609

\REF GW FGH \Gref
    C.-T. Gottstein, R.F. Werner
    "Ground states of the infinite q-deformed Heisenberg
    ferromagnet"
    Preprint Osnabr\"uck, 1994; archived as {\tt cond-mat/9501123}

\REF Isr Israel \Bref
    R.B. Israel
    "Convexity in the theory of lattice gases"
    Princeton University Press, Princeton 1979

\REF Mie Miek \Jref
    J. Mi\c ekisz
    "The global minimum of energy is not always a sum of local
    minima --- a note on frustration"
    J.Stat.Phys. @71(1993) 425--434

\REF PS Pasquier \Jref
    V. Pasquier, H. Saleur
    "Common structures between finite systems and conformal field
     theories through quantum groups"
    Nucl.Phys.B @330(1990)523--556

\REF RW RelEnt \Jref
        G.A. Raggio, R.F. Werner
        "Minimizing the relative entropy in a face"
        Lett.Math.Phys. @19(1990) 7--14

\REF Sew Sewell \Bref
       G.L. Sewell
       "Quantum theory of collective phenomena"
       Clarendon Press, Oxford 1986

\REF Vets Vets \Gref
   P. Vets
   "Boundary conditions for quantum systems as convex sets of states"
   Annex PhD thesis, Leuven 1990

\REF Mat Mathematica \Bref
   \noinitial. Wolfram Research{,} Inc.
   "Mathematica 2.2"
    Wolfram Research, Inc., Champaign, Illinois 1992

\vfill\eject
\let\Figure\onlycap
\bgsection . Figure Captions

\Figure C4C3\ad\Exa{C4C3}(@)
   state space of $\C(\set{1,2,3,4})$ with $\SAH$ embedded.
   The triangle is $\conv\SAHp=\SRE$. \\

\Figure Q2Q2\ad\Exa{Q2Q2}(@)
section of state space, with $\SAH$ marked ($\beta=3$, $\alpha=0.3$,
$\gamma=0.05$). The point on the vertical axis is $\omin^0$. It
is not contained in the convex hull of $\SAH$.\\

\Figure C3Q2a\ad\Exa{C3Q2}(@)
state space of $\M_2$ in logarithmic scale; inner circle is $\lntw\SAH$,
extension by curves is $\lntw\SRE(\M_2)$, angles determine $\lntw\SEE$. \\

\Figure C3Q2b\ad\Exa{C3Q2}(@)
upper left: the cones $C_1,C_2,C_3$ from \eq(SEEcones). Remaining
panels: $\lntw\SAH+ C_i$, $i=1,2,3$. The intersection of these sets
is shown in \Fig{C3Q2a}.\\

\Figure C3C2\ad\Exa{C3C2}(@)
left panel: state space of $\Cc3$; right panel: same figure in
logarithmic coordinates (same orientation). Thick line is $\SAH$,
lens shape is $\SRE$, and dashed corners together with $\SAH$ are
the boundary of $\SEE$.\\

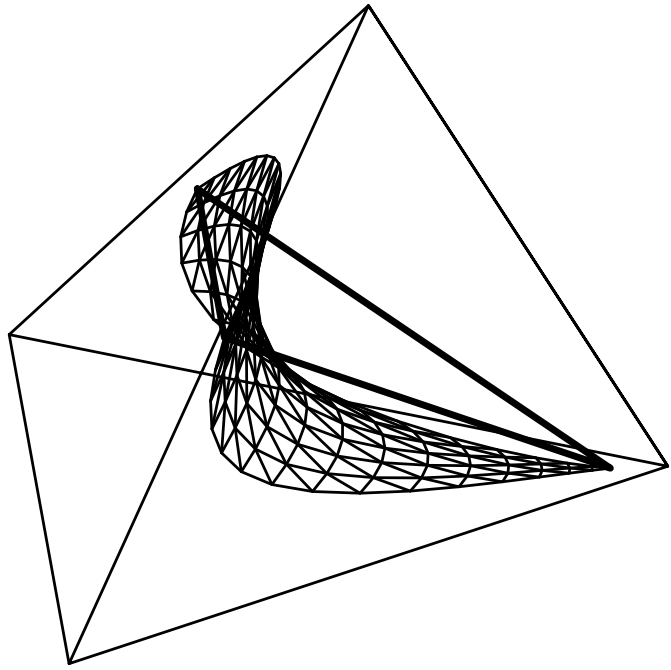
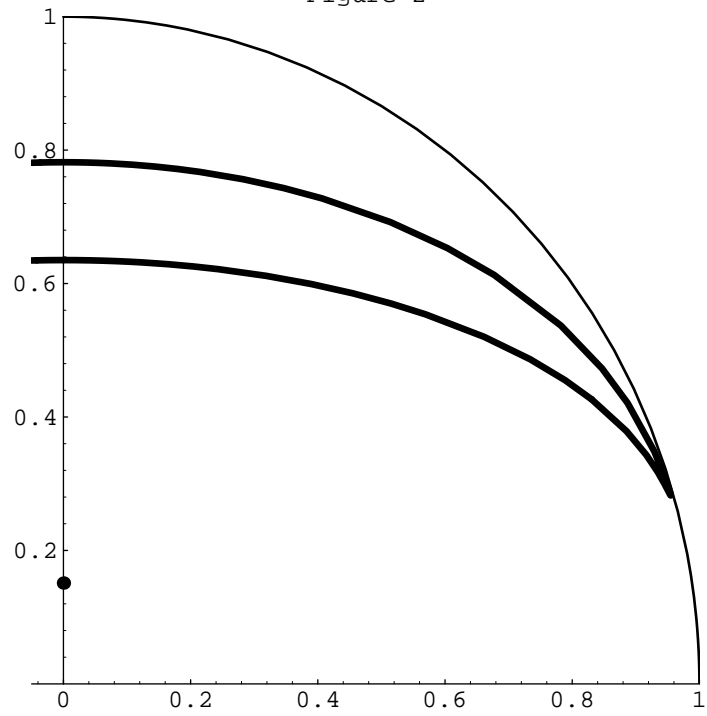
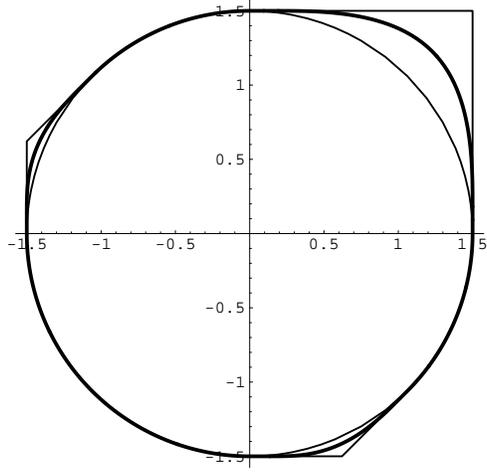
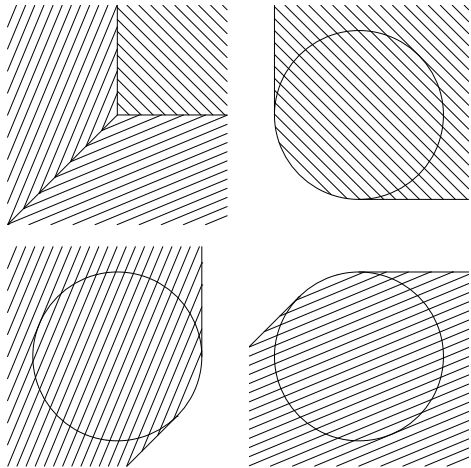
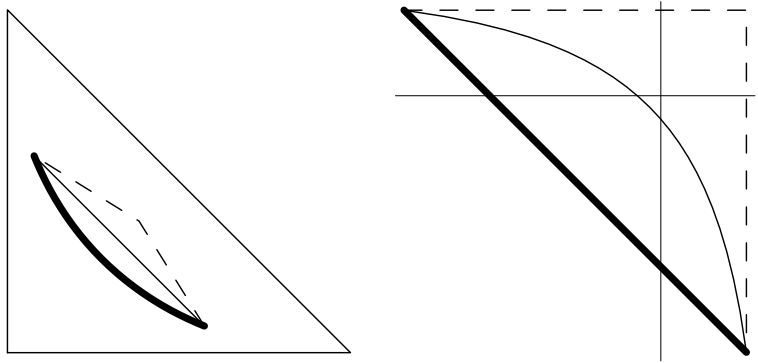
\bye